\documentclass[aps,prb,preprint,groupedaddress]{revtex4-1}

\usepackage{graphicx}

\usepackage{graphics}

\usepackage{color}

\usepackage{comment}

\begin{document}

%\preprint{}


\title{Creation of Localized Skyrmion Bubbles in Co/Pt Bilayers using a Spin Valve Nanopillar}


\author{Jennifer L. Grab}
\email{jlg373@cornell.edu}


%\thanks{}


\author{Alison E.~Rugar}

\author{Daniel C.~Ralph}
\email{dcr14@cornell.edu}
\altaffiliation{Kavli Institute for Nanoscale Science, Cornell University, Ithaca, NY 14853, USA}
\affiliation{Physics Department, Cornell University, Ithaca, NY 14853, USA}


\date{\today}

\begin{abstract}

We fabricate devices in which a magnetic nanopillar spin valve makes contact to a Co/Pt bilayer thin film with perpendicular magnetic anisotropy, in order to achieve local control of domains in the Co/Pt bilayer underneath the nanopillar.  The goal is to develop the ability to nucleate, detect, and annihilate magnetic skyrmions in the Co/Pt using spin-polarized currents from the nanopillar. We demonstrate the ability to distinguish the local behavior of the Co/Pt film beneath the nanopillar from the extended film and show that the two can switch independently of each other. This allows us to isolate a localized domain under the pillar that can be controlled separately from the rest of the Co/Pt film using applied currents and magnetic fields.  Micromagnetic simulations indicate that this localized domain has skyrmion symmetry. Our results represent a first step toward controlling room-temperature skyrmions using localized spin-transfer torque.

\end{abstract}



%\keywords{}


\maketitle




\section{Introduction}

Magnetic skyrmions, topologically-stabilized local spin textures, are of interest for their potential to enable highly dense, thermally stable information storage \cite{ReviewFert,NagaosaReview}. They can be as small as several nanometers in diameter and, since they can be moved readily via spin orbit torque, skyrmions have been proposed as candidates for racetrack memory \cite{track,Racetrack}.  Room-temperature skyrmions have been observed in material systems with an interfacial Dzyaloshinskii Moriya Interaction (DMI) \cite{dzyalo,moriya,FirstInterfacial,InterfacialTheory}, the most common example of which is the interface between a heavy (5$d$) metal (HM) with strong spin-orbit coupling and a ferromagnet (FM) with perpendicular magnetic anisotropy (PMA) \cite{PtCoMgO,PtCoIr,IrFeCoPt,Woo,CoFeBMgO,LTEM}.  Previous experimental work has shown that magnetically-soft HM/FM bilayers that host chiral stripe domains at zero field can support isolated skyrmions at a nonzero applied out-of-plane magnetic field below saturation \cite{CoPtDomains}.  As the field increases, the stripe domains pinch off and shrink into skyrmions.

To utilize skyrmions for technologies, it is necessary to develop techniques to reliably create, detect, move, and delete them at controllable positions. Techniques using current from a spin-polarized scanning tunneling microscope (SP-STM) tip \cite{STMwriting,STMEfield} or exposure by a focused ion beam \cite{FIB} have demonstrated a high degree of control in the creation of skyrmions at defined locations, but these techniques require specialized equipment and cannot be used in ambient conditions.  Creating skyrmions using lateral current flow through a geometric constriction, {\it i.e.} ``blowing skyrmion bubbles," works in ambient conditions, but these devices have size scales on the order of microns and do not have the capacity to read or delete skyrmions \cite{BlowingBubbles,Bubbles2,Bubbles3}.  In this paper, we report the first steps to investigate what is effectively the device equivalent of the SP-STM technique, by making multi-terminal devices in which a spin-valve nanopillar can apply a spin-polarized current locally to a HM/FM bilayer.  Creating skyrmions in an extended film using a spin-valve nanopillar is advantageous because nanopillars can be made with very small diameters, with the potential to create compact skyrmions.  Skyrmions can also be stabilized by interlayer coupling with the pillar, and the pillar can serve to read out the presence of skyrmions via giant magnetoresistance (GMR).  

The possibility to create skyrmions using a local spin-transfer torque in a nanocontact geometry has been analyzed previously using micromagnetic simulations. Sampaio {\it et al.} \cite{disktheory} considered an 80-nm diameter disk of Co/Pt with spin polarized current injected into a 40-nm circular region in the center.  For certain current densities and DMI values, a skyrmion could be created in the disk, and skyrmion formation could be assisted by applying a small perpendicular field or decreasing the PMA.  Similar results were also found for a 20 nm diameter pillar on a 80 nm square film by Kang {\it et al.} \cite{skyrmionelectronics}.   D\"{u}rrenfeld {\it et al.}\cite{nanocontact} modeled the injection of a large {\it unpolarized} current from a nanocontact into a Co/Pt layer, and found that the spin Hall effect associated with the lateral current flowing through the Pt might nucleate a skyrmion.  Sample geometries similar to ours have also been considered by micromagnetic simulations which analyzed the microwave-frequency dynamics of skyrmions in response to spin-polarized currents, for use in nano-oscillators and microwave detectors \cite{microwave,droplets,oscillator}.

\section{Film Growth and Device Fabrication}

Our focus is on domain manipulation in Co/Pt bilayers.  Skyrmions have been observed in sputtered Pt/Co/MgO films using photoemission electron microscopy combined with X-ray magnetic circular dichroism (XMCD-PEEM) \cite{PtCoMgO}.  First principle and \textit{ab initio} calculations for these films have shown that the Co/Pt interface has a large DMI even in the absence of MgO \cite{CoPtAnatomy,FertArxiv}.  

\begin{figure}
\includegraphics[width=\textwidth]{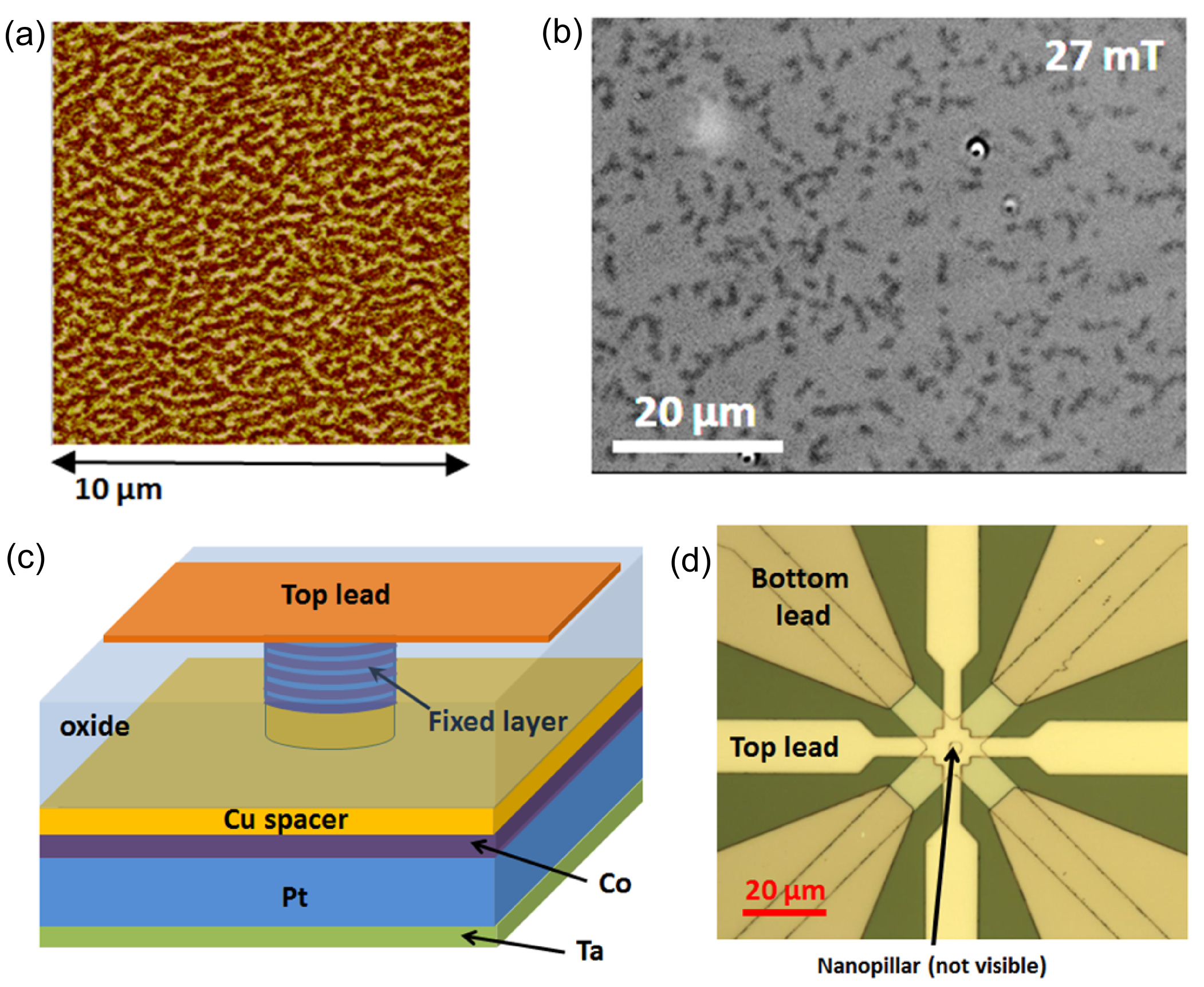}
\caption{\textbf{(a)} Magnetic force microscopy image showing stripe domains in the Co(1.65)/Pt(10) film used in this experiment.  \textbf{(b)} Widefield magneto-optical Kerr effect (MOKE) image of a similar film in an applied perpendicular field, showing isolated domains.  The film is initially saturated in the negative direction, which corresponds to the darker regions.  The magnetic field value is approximate. \textbf{(c)} Cartoon showing a side view of the spin-valve device geometry.  Features are not to scale.  \textbf{(d)} Optical microscopy image of a finished device with a 100-nm-diameter pillar.  The square in the center is a hole through the oxide layer to allow the top leads to contact the pillar.  The pillar is too small to see.}
\label{DeviceDesign}
\end{figure}

Our films were deposited using DC magnetron sputtering onto high-resistivity silicon substrates with a chamber base pressure $< 4 \times 10^{-9}$ Torr, and characterized using vibrating sample magnetometry and magnetic force microscopy.  The layer thicknesses for the platinum (10 nm) and cobalt (1.65 nm) were tuned such that the PMA is weak enough to favor the existence of stripe domains at zero field after saturation (Fig.~\ref{DeviceDesign}(a)).  For nonzero applied magnetic fields below saturation, we observe isolated domains that are likely skyrmion bubbles (Fig.~\ref{DeviceDesign}(b)) based on similar observations in the literature \cite{PtCoIr} and because these films have been shown to favor N\'{e}el walls \cite{Woo}.  Including the fixed-layer polarizer and spacer layer, our entire multilayer stack is substrate/Ta(3)/Pt(10)/Co(1.65)/Cu(5)/ [Co(0.4)/Pt(1.6)]$_8$/Co(0.4)/Pt(10), with thicknesses in nanometers.  The [Co(0.4)/Pt(1.6)]$_8$ multilayer that is used as the fixed-layer polarizer is chosen to have a large coercive field relative to the bilayer and a sharp hysteresis loop. Figure \ref{SoftGMR2}(a) shows the magnetization vs.\ applied magnetic field loop for the film, with the magnetic field oriented out-of-plane.  

We pattern this multilayer stack into the spin-valve-like device geometry shown in Fig.~\ref{DeviceDesign}(c,d).  The bottom leads of the device consist of the Co/Pt bilayer and Cu spacer patterned into a cross shape with channel widths ranging from 2 $\mu$m to 100 $\mu$m. The fixed layer is etched into a 100-nm-diameter nanopillar in the center of the cross.  The area immediately surrounding the pillar is covered in a protective oxide layer, and then top and bottom contacts are made such that spin-polarized current can be applied through the pillar into the bilayer underneath.  Simultaneously or separately, current can also be applied within the Co/Pt layer, passing beneath the nanopillar. 

We use a five-step lithography process to make these devices.  First, the full extended film stack is etched into Hall crosses using photolithography and Ar ion milling.  The pillars are then defined using electron beam lithography and ion milling, timed to etch partially through the Cu spacer.  We deposit 70 nm of protective SiO$_2$ using electron beam evaporation before resist removal.  Third, Ti/Pt bottom contacts are fabricated via photolithography, an RIE etch through the SiO$_2$ layer, and sputtering.  An extra 80 nm of protective SiO$_2$ is deposited around the pillar to prevent shorting between the top and bottom leads.  Finally, we fabricate Ti/Pt top contacts using photolithography and sputtering. 

GMR measurements through the nanopillar are used to monitor the state of the pillar and the Co/Pt bilayer region immediately underneath.  Currents applied through the nanopillar can also be used to exert a spin-transfer torque on the Co/Pt bilayer to reorient its magnetization.  We define positive current to correspond to electron flow up the pillar, which, via spin-transfer torque, favors antiparallel alignment of the Co/Pt free layer relative to the fixed layer. An additional feature of our device design is that, because the bottom leads are patterned into a cross shape, we can independently monitor the state of the extended Co/Pt bilayer using the anomalous Hall effect (AHE).  This allows us to distinguish behavior local to the pillar from effects in the extended film.  We make devices with a variety of Hall cross widths, which further helps to separate out bulk effects.  

\section{Isolation and Control of Localized Domains Under the Nanopillar}

\begin{figure}
\includegraphics[width=3.4in]{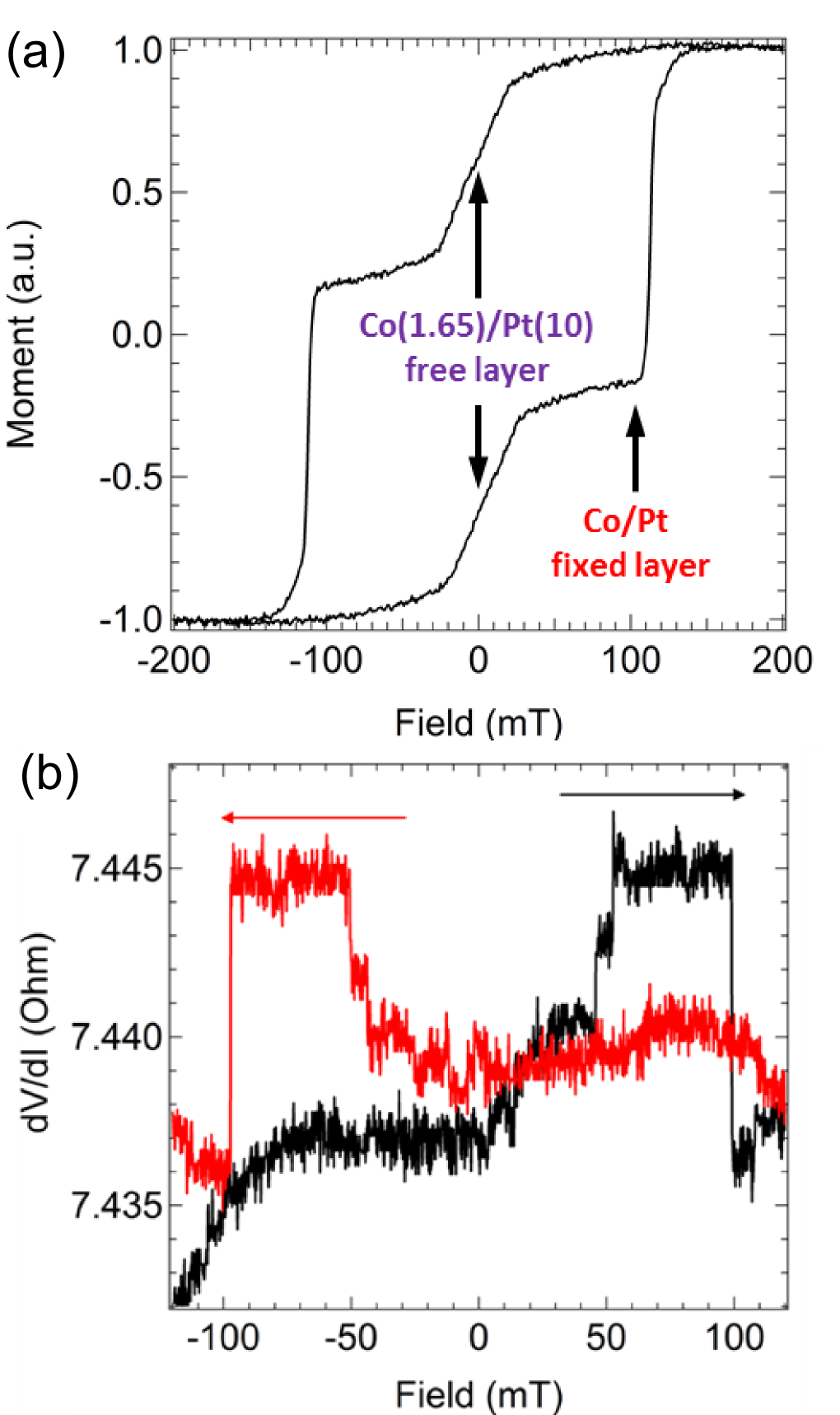}
\caption{\textbf{(a)}  Magnetometry data for the extended film before device fabrication.  \textbf{(b)} GMR measurement of a nanopillar device.}
\label{SoftGMR2}
\end{figure}

\begin{figure}
\includegraphics[width=\textwidth]{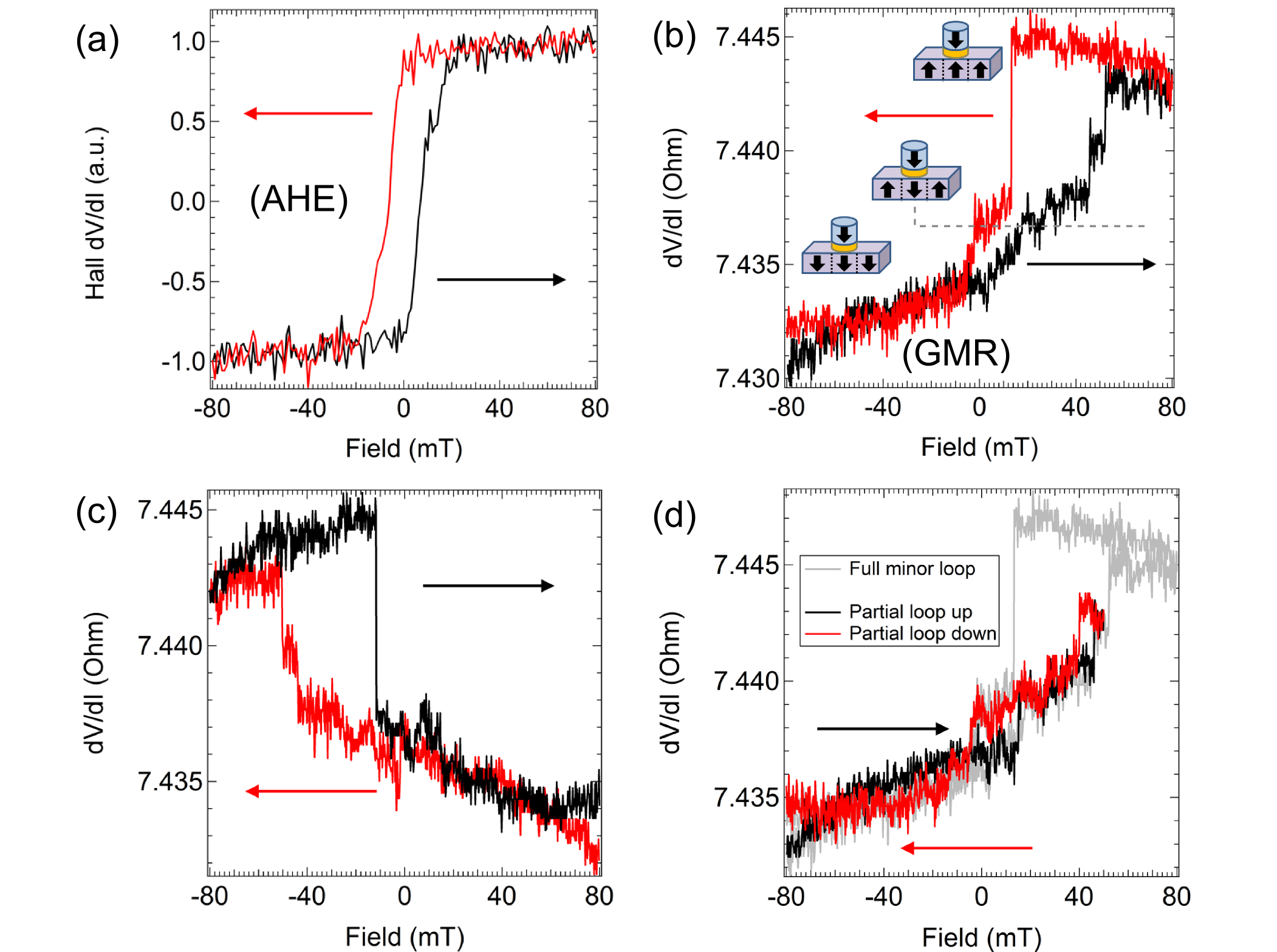}
\caption{\textbf{(a)} AHE data showing switching of the extended Co/Pt bilayer.  \textbf{(b)}  The GMR minor loop with the nanopillar fixed layer saturated in the negative field direction shows that the Co/Pt bilayer under the pillar switches at a much higher applied magnetic field than the rest of the same Co/Pt bilayer.  The insets show cartoons of the magnetic configuration at various points in the field sweep.  \textbf{(c)}  GMR minor loop with the nanopillar fixed layer saturated in the positive field direction.  \textbf{(d)}  A smaller field loop shows that the slope is repeatable and reversible.}
\label{softminor}
\end{figure}

Unless otherwise noted, all results discussed here are for the same device, for which the extended Co/Pt layer has a Hall cross geometry with 2 $\mu$m wide channels.  Similar behavior has been observed in six samples.  Four-point measurements of the nanopillar magnetoresistance show two clear switching events corresponding to the fixed and free layers, as expected (Fig.~\ref{SoftGMR2}(b)).  Slight misalignments in the pillar position cause a series resistance of several ohms.  Using the room-temperature  resistivity values  $\rho_{\text{Pt}}=21 \times 10^{-8}$ $\Omega \cdot$m, $\rho_{\text{Co}}=5.6 \times 10^{-8}$ $\Omega \cdot$m, and $\rho_{\text{Cu}}=1.7 \times 10^{-8}$ $\Omega \cdot$m, we calculate a GMR of approximately 1.5\%.  The AHE signal confirms that the lower-field switching event corresponds to the bilayer (Fig.~\ref{softminor}(a)).  However, interestingly, switching of the Co/Pt bilayer as measured by GMR occurs at a higher applied field compared to measurements of the AHE on the same device. This difference is even more striking in GMR measurements of the minor loop (Fig.~\ref{softminor}(b)), with the fixed-layer pillar magnetization pointing down.  The hysteresis curve for the Co/Pt bilayer, rather than being centered around zero field, is shifted, with the lower-resistance state more favorable at lower fields.  We ascribe this shift to a localized ferromagnetic interlayer coupling between the fixed-layer pillar and the Co/Pt bilayer, that causes the Co/Pt film underneath the nanopillar to switch at different magnetic fields than the extended Co/Pt bilayer.  We confirm this by measuring the reverse minor loop of the Co/Pt bilayer, with the nanopillar magnetization fixed pointing up (Fig.~\ref{softminor}(c)). In this case, the hysteresis loop is offset toward negative fields, again making parallel alignment more favorable at low fields.

If we perform a smaller field loop which switches and saturates the extended Co/Pt film but not the region under the pillar, the GMR signal becomes nonhysteretic, but has a repeatable slope (Fig.~\ref{softminor}(d)).  More interestingly, however, the AHE signal for the smaller loop still looks the same as Fig.~\ref{softminor}(a), meaning the extended Co/Pt film can be oriented opposite to the region of the Co/Pt film under the pillar.  These data therefore show that using the applied magnetic field alone, even in the absence of any applied current, we can localize a reversed domain under the pillar and control its lateral size.

\begin{figure}
\includegraphics[width=3.4in]{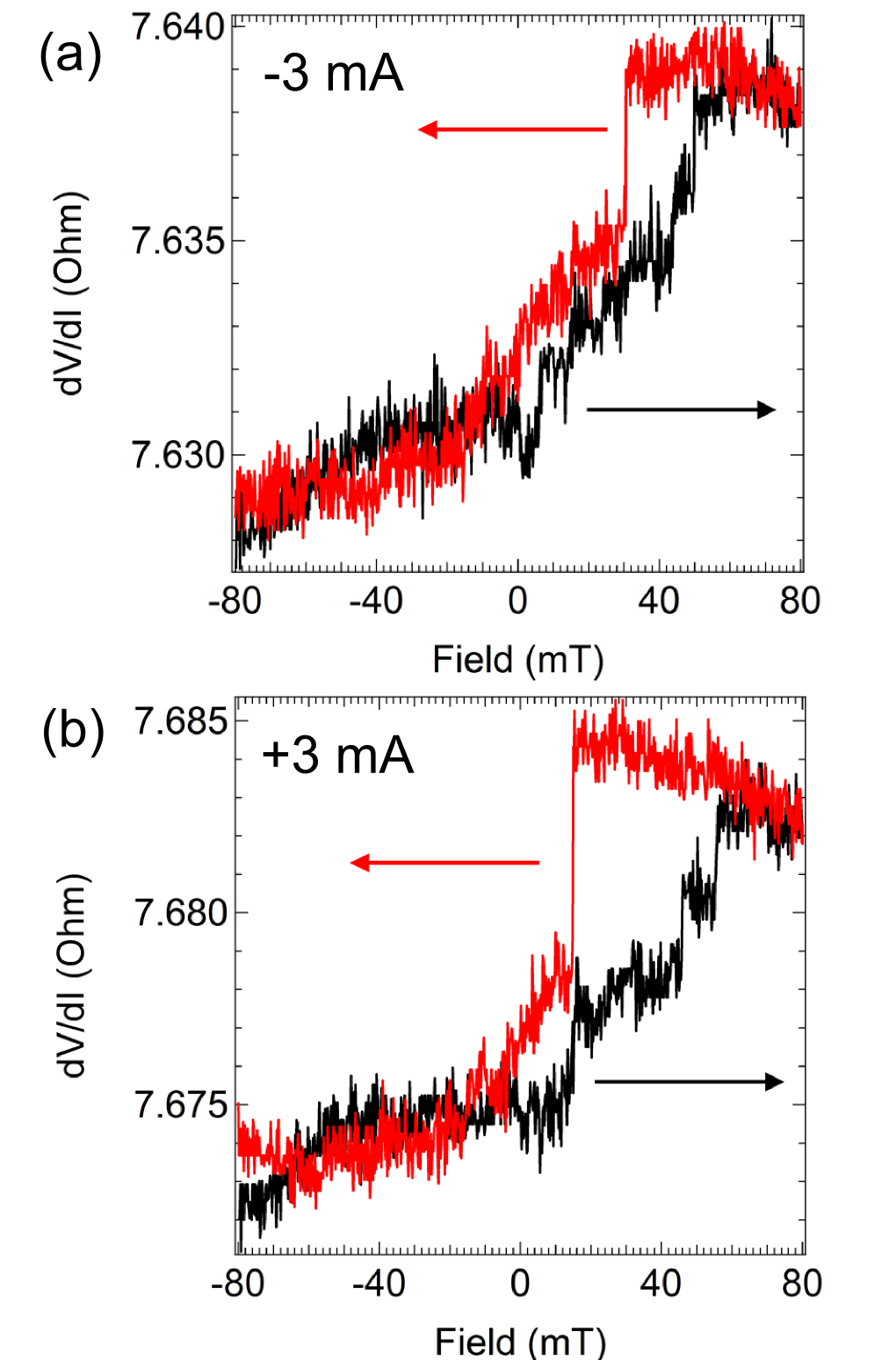}
\caption{Nanopillar GMR minor loop with the fixed-layer magnetization pointed down, measured with a constant current applied through the nanopillar of \textbf{(a)} -3 mA and \textbf{(b)} +3 mA.}
\label{fcur}
\end{figure}

By applying a spin-polarized electrical current through the nanopillar, we can control the value of applied magnetic field required to create the localized domain (Fig.~\ref{fcur}(a,b)).  Looking at the GMR minor loop, the switch from antiparallel to parallel (i.e., the switching event for a downward-sweeping magnetic field) is shifted depending on the sign of the applied current.  We interpret switching that is asymmetric with the sign of the current as a consequence of spin-transfer torque from the spin-polarized current. We find that a positive current stabilizes the antiparallel state, as expected from the sign of the spin-transfer torque. Both the applied magnetic field and the current through the nanopillar can therefore be used to manipulate the localized magnetic domain beneath the nanopillar. (We will analyze the phase diagrams as a function of current and magnetic field in more detail below.)  The AHE signal from the extended Co/Pt bilayer appears unchanged, regardless of applied current, as expected. The current seems only to affect the switch from antiparallel to parallel (i.e., creation of the localized domain) and not the reverse (erasure of the domain).  

\section{Comparison to Micromagnetic Simulations}

To help interpret the experimental results, we performed micromagnetic simulations using mumax3 \cite{mumax,mumaxweb}.  We used a mesh size of 4 $\times$ 4 $\times$  0.825 nm$^3$, which is much smaller than the domain length scale, and which allows the simulations to be 3-dimensional, i.e. there is more than one voxel in every direction.

From magnetometry data, we measured the saturation magnetization for Co(1.65)/Pt(10) films to be $M_s = 1.3 \times 10^6$ A/m, which is close to the bulk value for Co, 1.42 $\times 10^6$ A/m, and comparable to values for similar films in the literature \cite{PtCoPtPMA,PtCoAlDMI}.  Direct measurement of the first-order uniaxial anisotropy constant, $K_1$, for a perpendicular film is nontrivial if stripe domains are present \cite{Kooy,Kittel}.  Therefore, to estimate $K_1$ for our films we tested a range of reasonable values for both the DMI and $K_1$ in the simulations, looking for a combination that gives a domain size close to the value we observe experimentally ($\sim$ 100 nm using MFM).

\begin{figure}
\includegraphics[width=\textwidth]{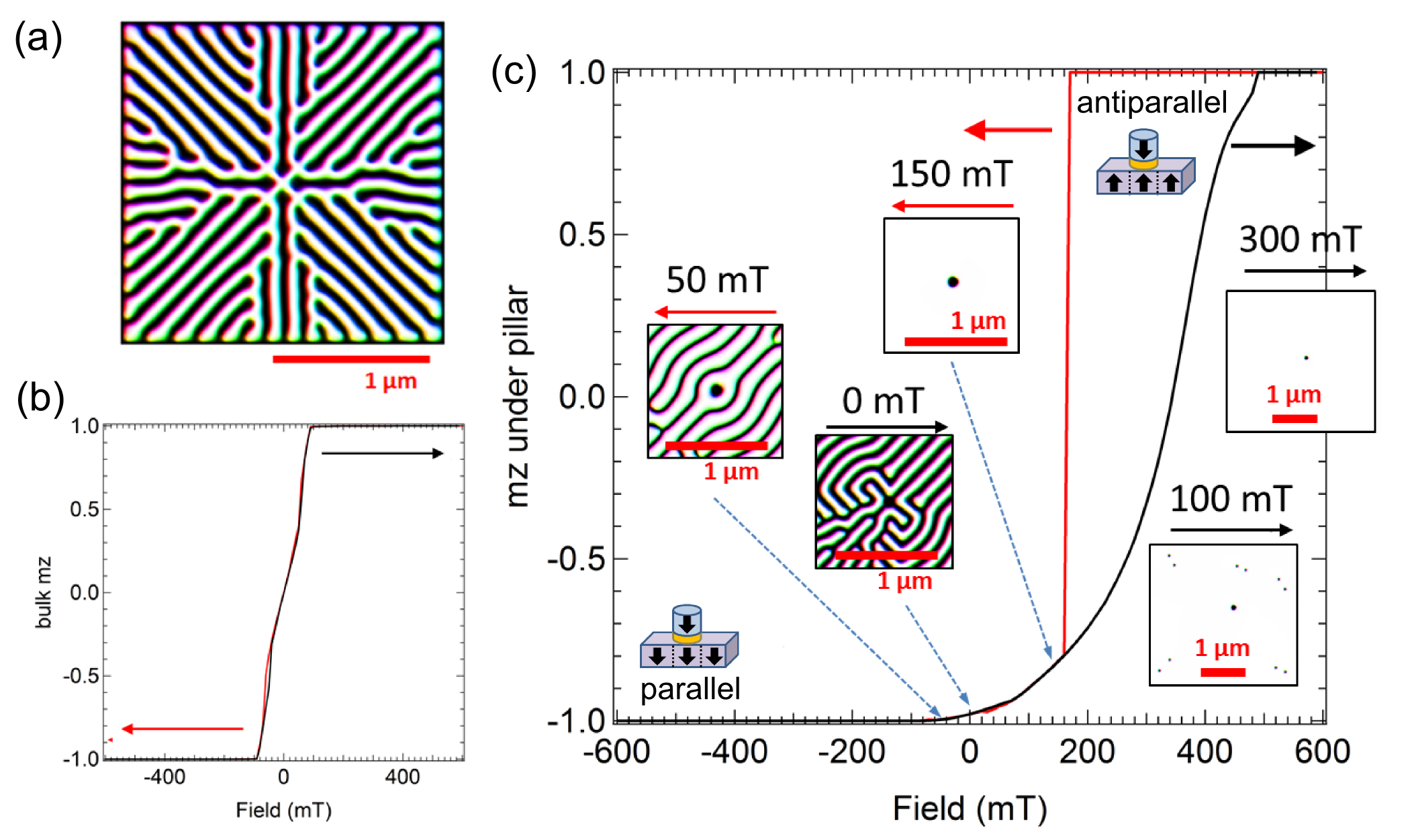}
\caption{\textbf{(a)} Simulated domain structure using parameters listed in text.  The symmetry is due to the assumed boundary conditions.  \textbf{(b-c)} Calculated average out-of-plane component of magnetization vs.\ field for the extended Co/Pt film and the region of film under the pillar, respectively.  The nanopillar fixed layer magnetization is oriented in the $-z$ direction (so that $m_z = -1$ corresponds to parallel alignment and $m_z = +1$ to antiparallel). The coupling field assumed is 300 mT.  The insets to \textbf{(c)} show snapshots of the domain structure at various points in the hysteresis curve.  White corresponds to moment oriented in the $+z$ direction, and the arrows indicate the direction of field change.  A N\'{e}el skyrmion is present under the pillar after the extended film has mostly saturated.  The skyrmion is erased when the field is increased beyond 500 mT.}
\label{mumax}
\end{figure}

We found that stripes exist in simulations following out-of-plane saturation only for a narrow range of anisotropy constant, $K_1 = 1.00 \pm 0.05 \times 10^6$ J/m$^3$.  For larger $K_1$, the film is uniform, and smaller $K_1$ results in an in-plane domain structure.  The range of DMI values, $D$, that support stripes is much larger.  The widest stripes are at $D = 1.5$ mJ/m$^2$ (Fig.~\ref{mumax}(a)), which is in agreement with previously measured and calculated values for the Co/Pt interface \cite{CoPtAnatomy,Woo,PtCoAlDMI}.  The simulated domain size using the postulated values for $K_1$ and $D$ is still somewhat smaller than that which we observed by MFM.  Since we have observed that the MFM tip often moves domains around, the imaged domains may be slightly broadened in the scanning direction.  Moreover, the simulations are done at T=0 K, and factors such as grain size and interfacial roughness were not taken into consideration.

We simulated devices with a 2 $\times$ 2 $\mu$m$^2$ Co/Pt bilayer and a 100 nm diameter pillar, which is emulated by defining a circular region within which there is a nonzero coupling field from the fixed layer and a spin-transfer torque due to the spin polarized current.  Only the Co/Pt extended layer is simulated, meaning that possible dynamics in the nanopillar fixed layer are not taken into account.  The pillar is slightly off center to eliminate effects from stripe domain symmetry, and for all simulations discussed here the pillar magnetization is assumed to be fixed in the $-z$ direction.  The sign convention for the current is the same as in the experiment.  The current densities we quote for applying spin-transfer torque should be understood as effective values $J' = J \times P$, where $J$ is the true current density and $P < 1$ is the spin polarization, since we do not independently determine the value of $P$.  We do not consider the effects of spin Hall torque generated by lateral current flow in the Cu/Co/Pt base layer\cite{nanocontact}. Compared to the simulations of D\"{u}rrenfeld et al.\cite{nanocontact}, the presence of the Cu layer and a thicker Pt layer should reduce this effect in our samples.

We calculated the minor hysteresis loop for devices with a range of different values for the ferromagnetic interlayer coupling field (always in the $-z$ direction), finding results closest to our measurements when the coupling field was greater than the coercive field of the Co/Pt bilayer (Fig.~\ref{mumax}(b,c)).  With no applied current, the effect of the interlayer coupling is to shift toward positive fields the magnetization vs.\ field hysteresis loop of the Co/Pt film under the nanopillar, making parallel alignment more favorable at low fields, whereas the extended film is unaffected.  For the case that the applied magnetic field is ramped from negative values to positive, above the coercive field of the Co/Pt extended layer there is a gradual transition for the localized region of Co/Pt underneath the pillar from parallel to antiparallel alignment (relative to the nanopillar fixed layer).  This change is due to the presence of a N\'{e}el skyrmion under the pillar, which shrinks in diameter as the magnitude of the field increases.  Due to the coupling field, this skyrmion can persist over a broader range of applied field than skyrmions occurring in the rest of the film.  When the applied magnetic field is swept in the opposite direction, from positive to negative, we find a much more abrupt switching event associated with the nucleation of a skyrmion (corresponding to a switch of the Co/Pt bilayer underneath the pillar from up to down, making it parallel to the fixed layer). The sharp transition is followed by a gradual slope to full parallel alignment. All of these features of the simulation are in good qualitative agreement with our experiment. We therefore conclude that the localized domain in the Co/Pt bilayer created underneath the nanopillar is a skyrmion bubble.

\begin{figure}
\includegraphics[width=3.4in]{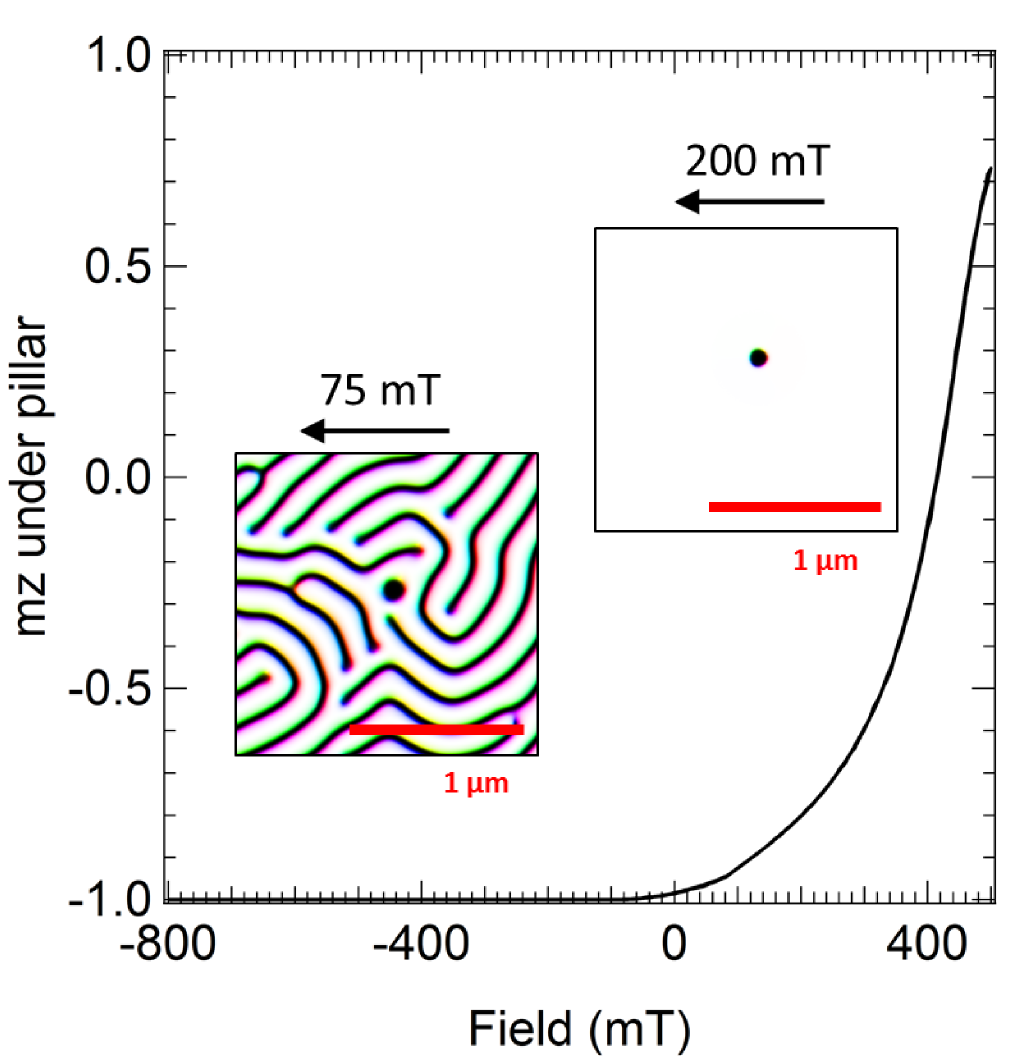}
\caption{Calculated $z$ component of the magnetization in the localized region of the Co/Pt bilayer under the nanopillar vs. applied magnetic field.  If the bilayer under the pillar doesn't fully switch to the $m_z = +1$ antiparallel state, the field dependence is nonhysteretic, and corresponds to a N\'{e}el skyrmion changing size under the pillar.  The insets show the domain structure after saturating the extended film (except the region under the nanopillar) in the $+z$ direction and ramping the field back down, while the nanopillar fixed layer magnetization always remains pointing in the $-z$ direction. }
\label{mumaxsub}
\end{figure}

If we simulate a magnetic-field loop such that the extended Co/Pt film switches, but not the region under the pillar, we see nonhysteretic behavior similar to our experimental results (Fig.~\ref{mumaxsub}).  Saturating the extended film (except for the region under the nanopillar) and then ramping the field back down, we observe a skyrmion bubble under the pillar that gets larger as the field magnitude decreases.  The skyrmion remains present even when stripe domains start to nucleate in the extended film.  

\begin{figure}
\includegraphics[width=3.4in]{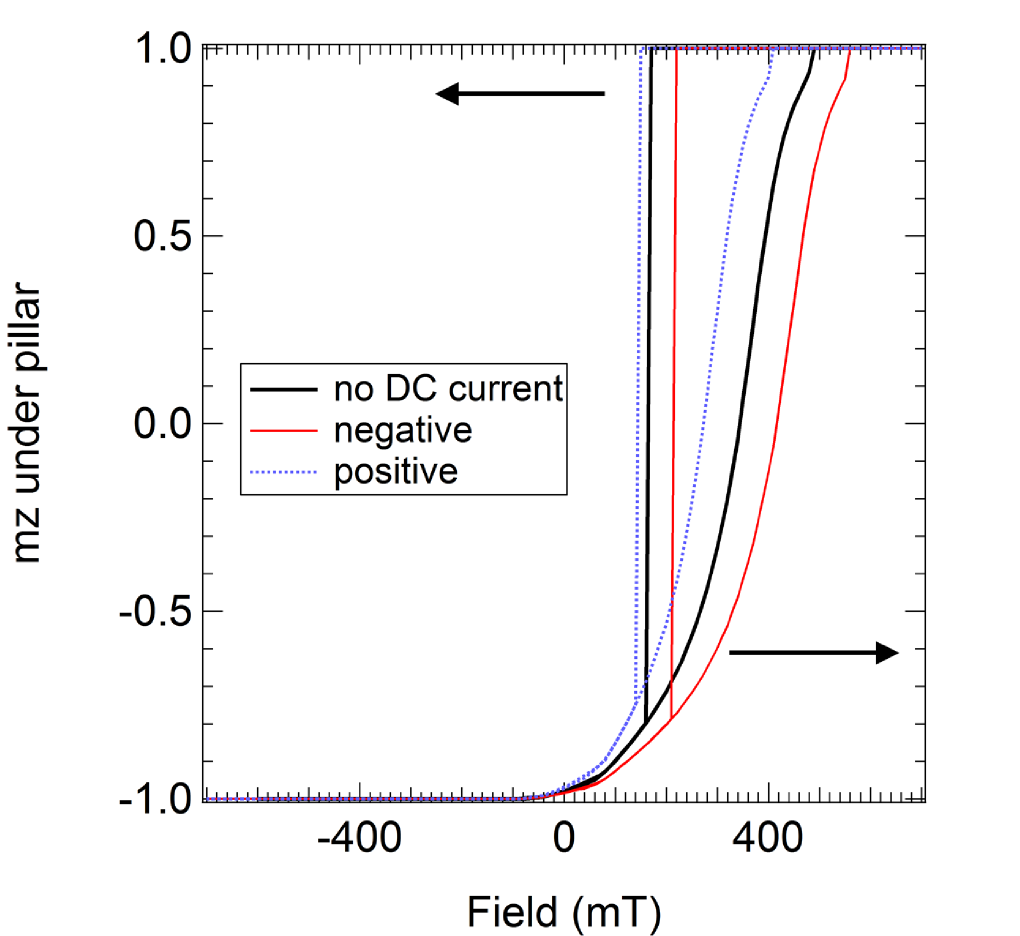}
\caption{Calculated z component of the localized Co/Pt bilayer magnetization under the nanopillar vs.\ field with different currents applied through the nanopillar.  The effective current density $J'$ is $\pm 5 \times 10^{7}$ A/cm$^2$ (an effective current of 4 mA through a 100-nm-diameter pillar) and the coupling field is 300 mT.}
\label{mumax2}
\end{figure}

When we apply spin-polarized current in the pillar region to exert a localized spin-transfer torque on the Co/Pt bilayer, the simulated hysteresis loop for the film under the pillar shifts to either higher (negative current) or lower (positive current) magnetic fields (Fig.~\ref{mumax2}).  For positive currents, antiparallel alignment is favored, and the skyrmion both annihilates and nucleates at lower fields.  On the other hand, a negative current stabilizes the skyrmion, which is parallel to the pillar, so that it persists for a larger field range and nucleates at a higher field as the field is swept down.  The effect of the spin torque on skyrmion creation in the simulations (the switch down from $m_z$=1 upon decreasing magnetic field) is similar to our measurements, but the simulations appear to differ from experiment for the upward field sweeps, where the simulations predict that the spin torque should affect the size of the skyrmion at a given field, but we detect no change within the noise level of the experiment. 

This discrepancy might be due to the fact that our simulations are done at T = 0 K, while our measurements are done at room temperature. In addition, spin Hall torque generated by current flowing laterally in the Pt film to exit the region under the nanopillar\cite{nanocontact} is not included in the simulations.  Mumax3 also does not include the Oersted field, but because it is mostly in-plane near the pillar and small relative to the applied magnetic field, we do not expect it to have a strong effect on the qualitative behavior.

Based on the dependence of the nanopillar resistance as a function of applied magnetic field, we can make a rough estimate of the size of the skyrmion bubble and how it varies with field.  Assuming the domain is circular,
\begin{equation}
d_{\mathrm{skyrmion}} = d_{\mathrm{pillar}} \sqrt{1-\frac{\Delta R}{\Delta R_{\mathrm {total}}}}
\end{equation}
where $d_{\mathrm{skyrmion}}$ and $d_{\mathrm{pillar}}$ are the diameters of the skyrmion and pillar, respectively.  $\Delta R$ is the change in the nanopillar GMR at a particular field from full parallel, and $\Delta R_{\mathrm {total}}$ is the total change in resistance from parallel to antiparallel.  The step-like nature of the gradual transition to antiparallel indicates that certain bubble diameters are more stable than others.  Except at large negative currents, the smallest diameter possible before the abrupt transition to antiparallel is about 40 nm, and this also appears to be the smallest size that can be nucleated using spin-transfer torque.

\section{Determination of the Experimental Phase Diagram}

\begin{figure}
\includegraphics[width=\textwidth]{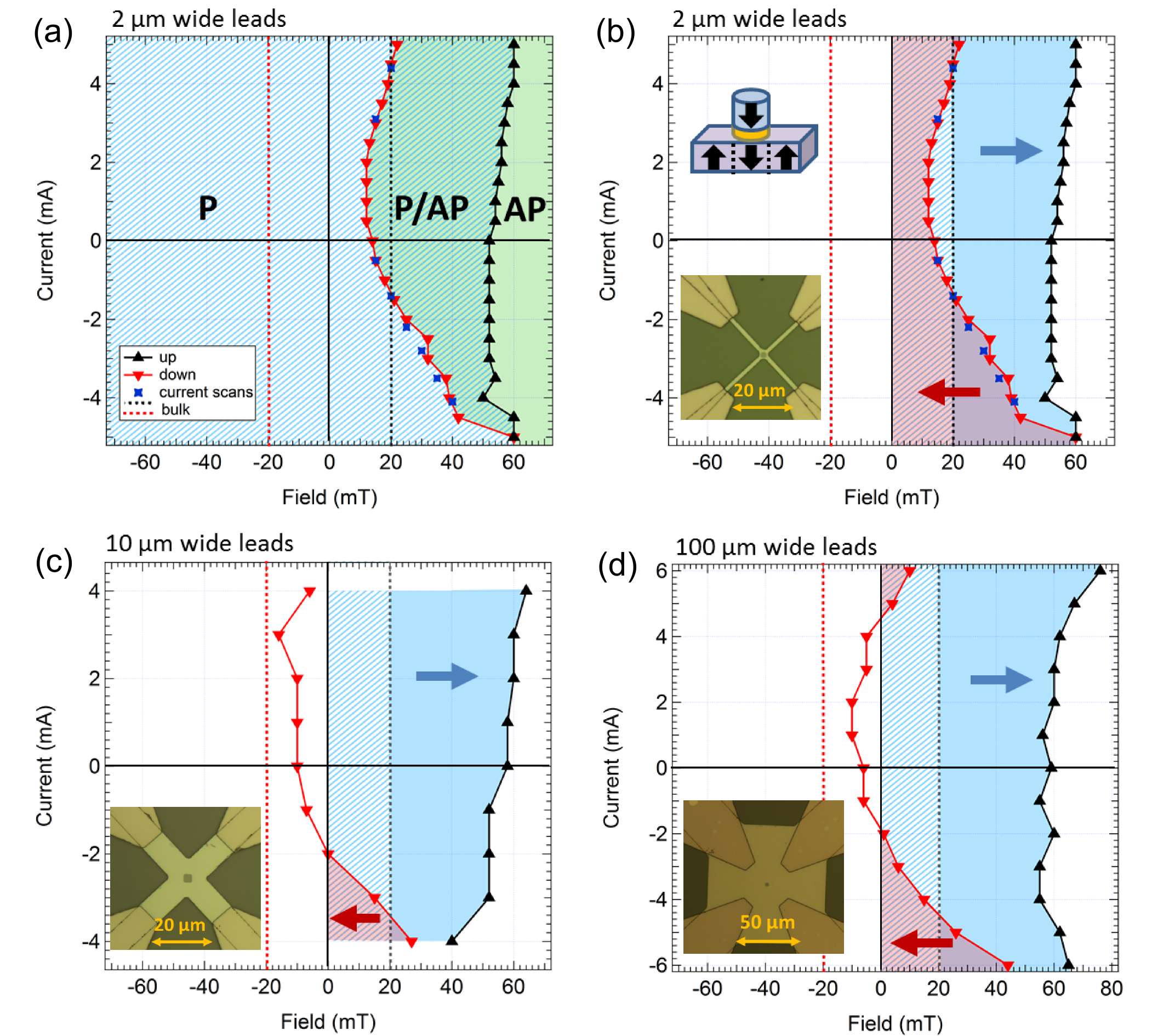}
\caption{\textbf{(a)} Phase diagram as a function of current through the nanopillar and out-of-plane applied magnetic field, for the device in which the extended Co/Pt layer is a Hall cross with 2 $\mu$m channels.  The black and red lines are determined from magnetic-field sweeps at fixed current as described in the text. Data points marked by the blue X's are determined from current sweeps at fixed field.  \textbf{(b)} Same phase diagram, recolored to show regions where an isolated skyrmion bubble may exist under the pillar (cartoon in inset). \textbf{(c-d)} Phase diagrams for devices where the extended Co/Pt layer is a Hall cross with 10 $\mu$m and 100 $\mu$m wide leads, respectively, with the stability regions for skyrmion bubbles marked using the same color coding as in (b). The insets show optical microscopy images of the device geometry before deposition of the top electrodes.}
\label{Phase}
\end{figure}

In order to better understand the dependence of current and magnetic field on the formation and annihilation of the skyrmion bubbles, we used GMR sweeps (e.g., Fig.~\ref{fcur}(a,b)) at different values of applied current to assemble experimental switching phase diagrams (Fig.~\ref{Phase}). Regions where the bilayer under the pillar is parallel (antiparallel) to the pillar are highlighted in striped blue (solid green).  The solid black line in Fig.~\ref{Phase}(a) marks where the bilayer under the pillar is finished switching up to be fully antiparallel (AP) to the nanopillar fixed layer.  This boundary seems to be mostly unaffected by a current applied through the nanopillar. The solid red line in Fig.~\ref{Phase}(a) marks the position of the abrupt switch down in the localized Co/Pt magnetization under the nanopillar, from the fully AP state to a state that is partially, but generally not fully, aligned with the fixed-layer magnetization.  This boundary has a strong dependence on applied current, which near zero current is clearly asymmetric in that positive current favors the AP state while negative current favors P. At larger magnitudes of applied current this boundary bends so that eventually both signs of current favor the P state.  This can be understood as an effect of heating reducing the coercivity of the Co/Pt layer in the presence of the ferromagnetic interlayer interaction with the nanopillar fixed layer. 

\begin{figure}
\includegraphics[width=\textwidth]{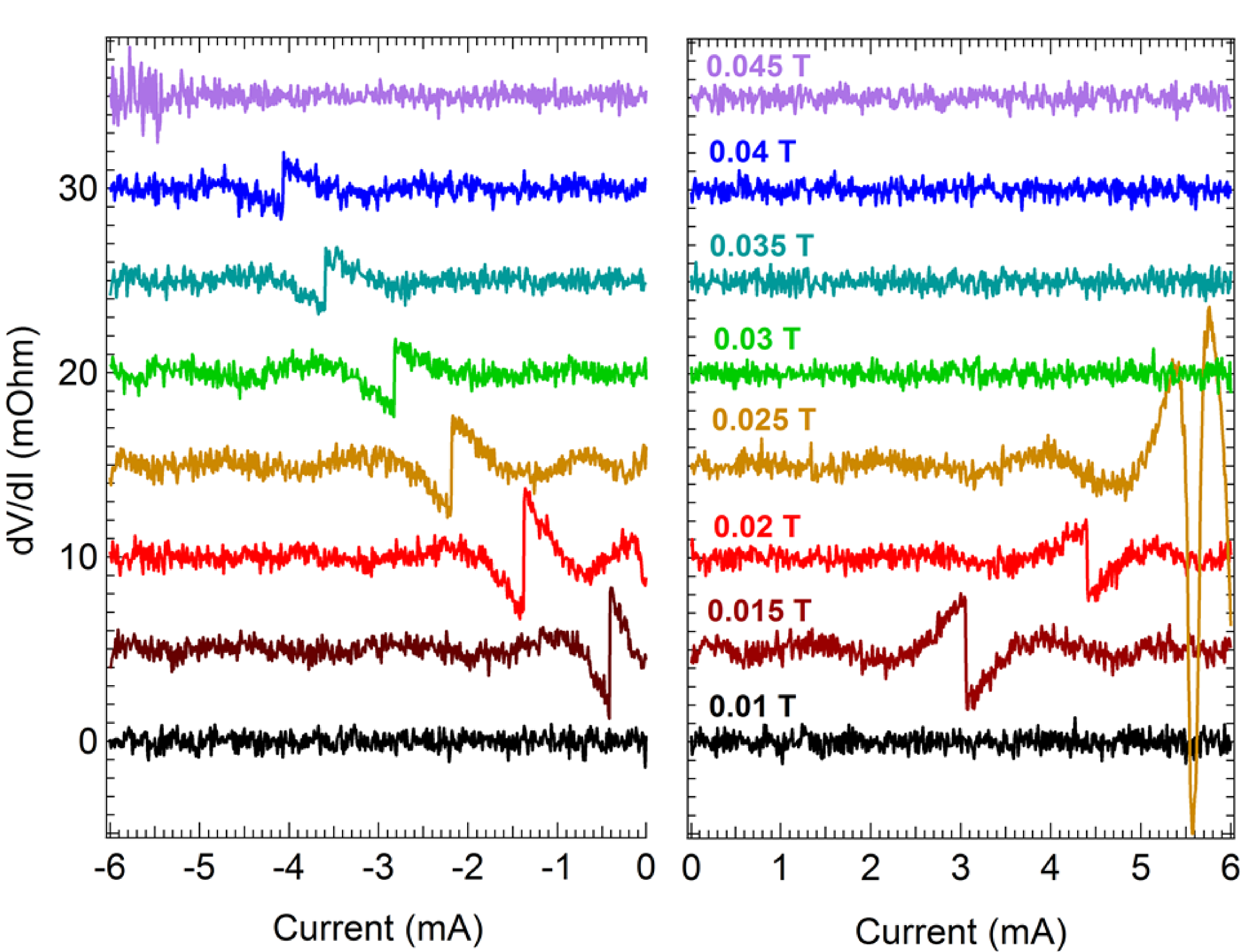}
\caption{Differential resistance of the nanopillar vs.\ applied current through the nanopillar at fixed values of magnetic field applied out-of-plane, showing current-induced resistance transitions. A parabolic background due to Joule heating is subtracted, and the data are offset for clarity.  Prior to each scan the Co/Pt bilayer is initialized to be antiparallel to the nanopillar fixed layer. The transitions observed are recorded as blue X symbols in Fig.~\ref{Phase}(a).}
\label{current}
\end{figure}

We checked this switching phase diagram by also performing scans with a swept current applied through the nanopillar, at fixed values of magnetic field.  Two sets of measurements were done, with the Co/Pt bilayer initialized either parallel ($m_z = -1$) or antiparallel ($m_z = +1$) to the nanopillar fixed layer. For current scans with the Co/Pt bilayer under the pillar initially in the parallel state, we observed switching only for applied magnetic fields near the coercive field, and these events were symmetric with the sign of current, indicating a dominant thermal effect. However, when the bilayer under the pillar was initially in the antiparallel state, we observed current-induced transitions (Fig.~\ref{current}) with locations in good general agreement with the swept-field transitions (blue markers in Fig.~\ref{Phase}(a)).  These switching events therefore correspond to current-induced nucleation of skyrmion bubbles beneath the nanopillar. These events are not observable in the AHE signal, confirming that that they can be identified as localized to the region of the Co/Pt bilayer under the nanopillar. As expected from the phase diagram, none of these switching events can be reversed with current alone, meaning that we do not observe current-induced annihilation of the skyrmion bubbles.  

For a full understanding of the region in which a skyrmion bubble can be stabilized, one must also consider the state of the extended Co/Pt bilayer in relation to the switching phase diagram, because to host a skyrmion bubble that is parallel to the nanopillar fixed layer the extended layer must be antiparallel to the fixed layer (i.e., extended layer pointed in the $+z$ direction).  The coercive field of the extended layer, as determined by the AHE, is about 20 mT, regardless of applied current.  The dotted black (red) lines in Figure \ref{Phase}(a) mark where the extended layer is done switching to the AP (P) state. When sweeping the external magnetic field up, a skyrmion bubble may exist at fields more positive than the coercive field of the extended Co/Pt layer, but less than the coercive field of the region under the pillar. This region is highlighted in solid blue in Fig.~\ref{Phase}(b).  When sweeping the external magnetic field down, a skyrmion bubble can exist for fields in between the abrupt switch of the Co/Pt region beneath the nanopillar away from the AP state and a field near zero where stripe domains begin to form in the extended film (the solid red region of Fig.~\ref{Phase}(b)).  The striped blue region of Fig.~\ref{Phase}(b) can also stabilize a skyrmion bubble if the external field is ramped up to a sufficiently positive value to saturate the extended Co/Pt layer but not so high as to switch the region under the nanopillar to the AP state, and then the external field is ramped back down. An isolated domain oriented antiparallel to the fixed layer is not possible in this system.

We have constructed similar diagrams for other devices in which the extended Co/Pt layer is patterned into a Hall cross made from channels with different widths, 10 $\mu$m (Fig.~\ref{Phase}(c)) and 100 $\mu$m (Fig.~\ref{Phase}(d)) wide leads. We anticipated that changing the geometry of the extended Co/Pt bilayer while keeping the pillar size the same should have little effect on the behavior of the Co/Pt bilayer under the pillar, and indeed the result for all three geometries is qualitatively similar. In devices with wider channels, however, a somewhat larger current is required to get the same effect from spin-transfer torque, and thermal effects seem slightly reduced.  It is possible these two effects are linked, in that the spin-transfer-induced switching is likely thermally assisted. 

\begin{figure}
\includegraphics[width=8.6cm]{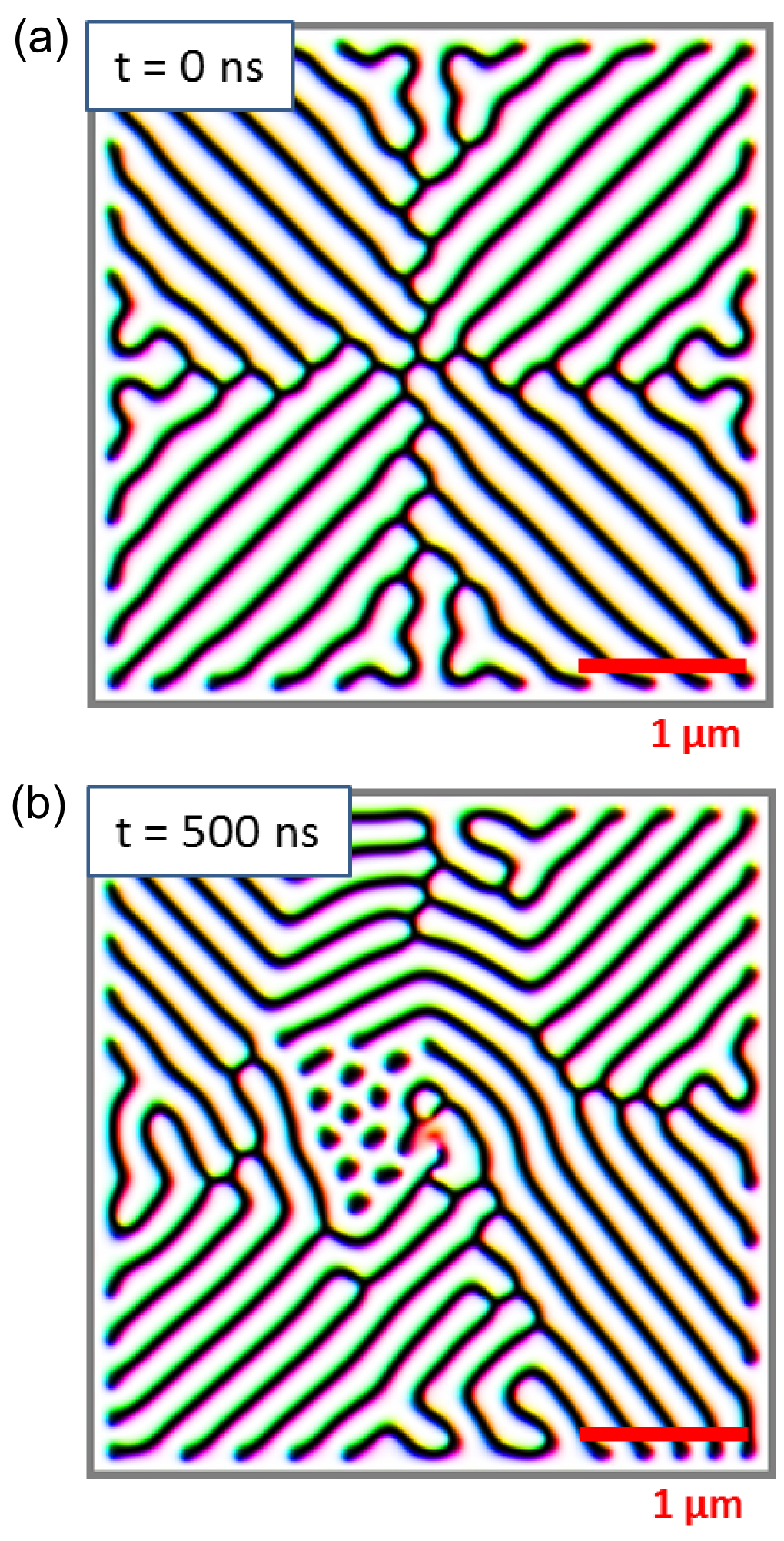}
\caption{Simulation of device with an in plane polarizer and an applied field of 60 mT.  Here, the interfacial DMI is $D$ = 2.0 mJ/m$^2$ and the effective current density is $J' = 5 \times 10^7$ A/cm$^2$. The 100 nm pillar is oriented in the $+x$ direction in the center of the film. Skyrmion bubble creation is observed at fields for which stripe domains exist in the vicinity of the pillar.}
\label{inplane}tcat
\end{figure}

\section{Simulations for an In-Plane-Polarized Nanopillar Fixed Layer}

As a potential idea for future work, we also simulated devices with an in-plane-magnetized nanopillar fixed layer, finding an interesting effect.  The device geometry continues to employ an out-of-plane magnetized Co/Pt extended layer with strong DMI and weak PMA, but we assume that the pillar magnetization is oriented in the $+x$ direction and that there is no longer any significant coupling between the fixed layer and the extended layer.  With the application of moderate current densities, the simulations indicate that out-of-plane domains under the pillar are pushed away by the in-plane-polarized spin current and they can sometimes pinch off repeatedly to form localized collections of skyrmion bubbles (Fig.~\ref{inplane}).  This effect occurs at fields where the Co/Pt extended film has stripe domains in the vicinity of the pillar, since new out-of-plane domains cannot be nucleated with this configuration.  This could potentially be a strategy to create controlled clusters of skyrmions.

\section{Discussion and Outlook}

We have demonstrated control of an isolated domain in a soft Co/Pt bilayer underneath a nanopillar fixed layer, using a combination of an interlayer interaction, spin-transfer torque, and an applied magnetic field.  Magnetoresistance measurements are in good qualitative agreement with simulations which indicate that we have achieved nucleation and control over a N\'{e}el skyrmion bubble.  

If we start in a state with the nanopillar fixed layer magnetization oriented opposite to a small applied out-of-plane magnetic field, and with the low-coercivity Co/Pt extended layer saturated parallel to the field, the presence of the skyrmion bubble under the nanopillar is indicated by a gradual, nonhysteretic change in the nanopillar resistance as a function of changing magnetic field.  This corresponds to a domain whose size decreases as the applied field increases.  At around 50 mT the skyrmion bubble annihilates, which is irreversible. Then, starting with an applied magnetic field strong enough to saturate the Co/Pt extended layer fully antiparallel to the nanopillar fixed layer, if the field is swept down we observe a sudden switch down in the nanopillar resistance that corresponds to nucleation of the skyrmion bubble, with an additional decrease in resistance due to its expansion as the applied field is reduced further. The field value required for nucleation of the skyrmion bubble can be shifted by an applied current through the nanopillar, with a sign consistent with expectations for spin-transfer torque from the current.  With an appropriately chosen bias field, an applied current can drive the nucleation of the skyrmion bubble, but we have not yet observed the opposite process of current-driven domain annihilation. 

This work represents a first step toward experiments which should allow us to explore and control the properties of skyrmions in several new ways.  We hope to be able to use spin-transfer torque from current applied through the nanopillar to drive high frequency dynamics of a skyrmion bubble trapped under the nanopillar, and explore their potential applications as nano-oscillators and microwave detectors \cite{microwave, oscillator, droplets}.  To date we have not measured any microwave emission in response to applied direct current in bilayers with thinner cobalt and stronger PMA, as predicted by Carpentieri, {\it et al.} \cite{droplets}; this may require optimization of the nanopillar fixed layer materials to achieve a higher spin polarization.  We will investigate whether it is possible to use spin-orbit torques from currents applied within the Co/Pt extended layer to take a skyrmion bubble nucleated controllably under one nanopillar and translate it elsewhere in the sample \cite{disktheory,BlowingBubbles}, and possibly detect it by shifting it under a different nanopillar.  The creation of skyrmion lattices that are electrically controllable may be possible by fabricating arrays of nanopillars.  We will also be working to integrate our devices with advanced imaging techniques \cite{LTEM, SQUID}, so that it might be possible to make direct observations of the skyrmion bubbles as we create, manipulate, move, and delete them with increasing levels of control.

\begin{acknowledgments}
We thank D. MacNeill, M. Guimaraes, C.-F. Pai, J. Park, and R. Buhrman for discussions and assistance with experiments, and we thank C. Garg for MOKE imaging.  This research was supported by the Cornell Center for Materials Research with funding from the NSF MRSEC program (DMR-1120296,DMR-1719875). This work was performed in part at the Cornell NanoScale Facility, a member of the National Nanotechnology Coordinated Infrastructure, which is supported by the NSF (ECCS-1542081).
\end{acknowledgments}


\bibliography{CoPtPaper_references}

\end{document}